# Truly Sub-Nyquist Method Based Matrix Pencil and CRT with Super Resolution

Huiguang Zhang[1], Baoguo Liu[2]

## Abstract

The emergence of ultra-wideband (UWB) and high-throughput signals has necessitated advancements in data sampling technologies1. Sub-Nyquist sampling methods, such as the modulated wideband converter (MWC) and compressed auto-correlation spectrum sensing (CCS), address the limitations of traditional analog-to-digital converters (ADCs) by capturing signals below the Nyquist rate. However, these methods face challenges like spectral leakage and complex hardware requirements. This paper proposes a novel super-resolution generalized eigenvalue method that integrates the matrix pencil method with the Chinese Remainder Theorem (CRT) to enhance signal processing capabilities within a true sub-Nyquist framework3. This approach aims to improve frequency resolution and accuracy in high-frequency signal extraction, with potential applications in telecommunications, radar, and medical imaging.

## 1.Introduction

The emergence of ultra-wideband (UWB) and high-throughput signals has created an urgent demand for innovative advancements in data sampling technologies. In particular, sub-Nyquist sampling methods, such as the modulated wideband converter (MWC) and compressed auto-correlation spectrum sensing (CCS), are becoming increasingly relevant. These advanced sampling techniques are designed to capture signals at rates that fall below the traditional Nyquist rate, effectively addressing the limitations posed by high-frequency signals that often surpass the capabilities of conventional analog-to-digital converters (ADCs). Despite their potential, these sub-Nyquist methods encounter several significant challenges. One of the primary issues is the reduction in accuracy that can result from spectral leakage, which can distort the true representation of the sampled signal. Additionally, the complexities associated with the hardware implementation of these techniques can be daunting, as they frequently necessitate the use of specialized pseudo-random modulation devices. This requirement for specific hardware can complicate the deployment of these technologies in practical applications.
 Current subspace methods, such as the Multiple Signal Classification (MUSIC) algorithm and Matrix pencil, do offer some advantages in high-frequency signal extraction. However, they are still largely confined to the traditional Nyquist sampling range, which limits their effectiveness in accurately determining both frequency and amplitude of the signals. The inherent constraints of these methods highlight the need for new approaches that can transcend the limitations of existing technologies. In response to these challenges, a novel super-resolution generalized eigenvalue method has been proposed. This innovative technique seeks to integrate the matrix bundle method with the principles of the Chinese remainder theorem to enhance the capabilities of signal processing within a true sub-Nyquist framework. By leveraging random modulation and specialized filtering techniques, this new approach aims to significantly improve frequency resolution, thereby offering a more effective solution to the complexities associated with modern signal processing. This advancement holds the promise of not only enhancing the accuracy of high-frequency signal extraction but also expanding the potential applications of sub-Nyquist sampling technologies in various fields, including telecommunications, radar, and medical imaging.Furthermore, the integration of these methodologies could lead to breakthroughs in real-time data analysis, enabling faster decision-making processes in critical applications.

## 2.Preliminary

### 2.1.Chinese Remainder Theorem

The Chinese Remainder Theorem (CRT) is a fundamental result in number theory and algebra, which provides a method for solving systems of simultaneous congruences with pairwise relatively prime moduli. It states that if one has several congruences, each with a different modulus, and these moduli are pairwise coprime, then there exists a unique solution modulo the product of these moduli. This theorem has been extended and applied in various mathematical and computational fields, including error correction, cryptography, and algorithm design.

The CRT is one of the oldest theorems in mathematics, originally used for solving congruence equations in ancient China [1][2],It states that a system of linear congruences with pairwise relatively prime moduli has a unique solution modulo the product of its moduli[3],The theorem can be generalized to arbitrary algebraic structures using Universal Algebra, where it involves congruences on algebras [4],The CRT is crucial in designing efficient multi-modular algorithms, particularly in integer matrix multiplication [5][6].

It is used in error detection and correction in redundant residue number systems, with novel algorithms simplifying the correction process (Tor et al., 2004)The theorem has been extended to polynomial forms, group theory, and unitary rings, showing its versatility across different algebraic systems [7].The complexity of deciding whether a tuple of congruences forms a Chinese Remainder tuple is coNP-complete, but tractable for certain algebraic classes like vector spaces and distributive lattices[3].Fast algorithms leveraging the CRT can achieve significant computational gains, such as an asymptotic factor improvement in complexity through techniques like FFT-trading [7]. An example of CRT application is reconstructing integers from their residues modulo a set of coprime moduli, such as reconstructing a number from residues modulo 2 and 5 [8]. The CRT has been applied to Fibonacci numbers, demonstrating its utility in various mathematical problems [9].While the Chinese Remainder Theorem is a powerful tool in number theory and algebra, its application is not without challenges. The complexity of certain problems involving CRT, such as determining Chinese Remainder tuples, can be computationally intensive, though tractable in specific cases. Additionally, the choice of moduli and the structure of the algebraic system can significantly impact the efficiency and feasibility of CRT-based solutions [3].

### 2.2 . Matrix pencil method

The matrix pencil method, also known as the generalized eigenvalue algorithm, is a mathematical approach used to solve generalized eigenvalue problems, particularly those involving matrix pencils, which are pairs of matrices. This method is crucial in various applications, including signal processing, control theory, and structural analysis, due to its ability to handle complex and large-scale problems efficiently. The matrix pencil method reformulates problems into a generalized eigenvalue problem, allowing for the separation of linear and nonlinear parameters, which enhances computational stability and efficiency [10].Singular matrix pencils present challenges due to the discontinuity of eigenvalues. Traditional methods involve extracting the regular part using the staircase form and applying solvers like the QZ algorithm [11][12].

Recent approaches involve transforming singular pencils into regular ones using randomized modifications, which maintain the finite eigenvalues and improve numerical stability[11] [13].The multiparameter matrix pencil problem (MPP) generalizes the one-parameter problem by involving multiple complex matrices and scalars, requiring the matrix pencil to lose column rank [14].Solutions involve transforming the problem into simultaneous one-parameter MPPs using Kronecker commutator operators, simplifying numerical solutions [15].Randomized algorithms, such as the divide-and-conquer eigen solver, offer high probability success for well-behaved matrix pencils by regularizing pseudospectra through perturbation and scaling.These methods provide guarantees for parallel solvers and establish nearly matrix multiplication time as an upper bound for complexity[16].

In signal processing, the matrix pencil method is used for estimating generalized eigenvectors, crucial for applications like beamforming. Novel algorithms have been developed that do not require learning rates, enhancing real-world applicability [12].The method is also applied in ultrasonic guided wave technology for bone characterization, where it computes modal wavenumber and attenuation coefficient, demonstrating improved convergence and noise reduction [11].

While the matrix pencil method is highly effective in solving generalized eigenvalue problems, it is not without limitations. The complexity of the method can increase significantly with the size and condition of the matrices involved, potentially leading to computational inefficiencies. Additionally, while randomized methods improve stability, they may not always guarantee exact solutions, necessitating further refinement and validation in practical applications.

## 3.Theory

Proof

for a multifrequecy signal x(t) which can be denoted as

$$x(t) = \sum_{i=1}^{m} a_i\, e^{j(2\pi f_i t + \varphi_{i0})}. \tag{1}$$

then two sub-Nyquist samplings with different sampling frequencies $f_{s1},f_{s2}$ are performed on the x(t), the sampled sequency can de denoted as $x_1(t)$ and $x_2(t)$ respectively.

$$x_1(t) = \sum_{i=1}^{m} a_i\, e^{j(2\pi f_{1i} t + \varphi_{i0})}. \tag{2}$$

$$x_2(t) = \sum_{i=1}^{m} a_i\, e^{j(2\pi f_{2i} t + \varphi_{i0})} \tag{3}$$

Then, two hankel matrix $X_{11}$ and $X_{12}$ are constructed via $x_1(t)$, and another two hankel matrix $X_{21}$ and $X_{22}$ via $x_2(t)$ with time delay. From the following equations,(4), we can get two result matrix $R_1$ and $R_2$,which include the aliased frequency, magnitude, and phase, by the excellent work of Liu [17].

$$X_{11} = \lambda_1 X_{12},\; X_{21} = \lambda_2 X_{22}. \tag{4}$$

And the R1,R2 are illustrated as follows:

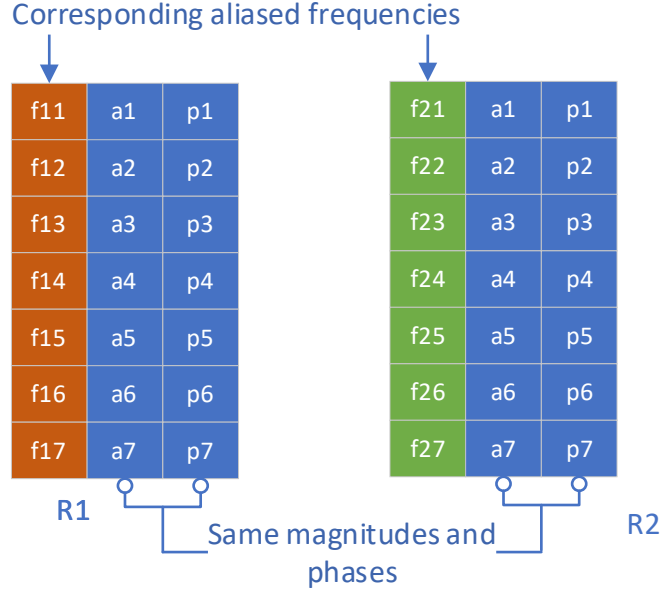

Figure 1  Same magnitudes and phases and corresponding aliased frequencies
Two matrix pencils set constructed from two different sampling frequencies yields the same amplitude and phase, but the frequencies are different.

We can easily select the corresponding aliased frequencies via their magnitudes and phases. then with the following aliased equations and CRT , we can get the real frequencies $f_{ri}$ easily.

$$f_{ri} = k_{1i}f_{s1} + f_{1i}, f_{ri} = k_{2i}f_{s2} + f_{2i}. \qquad (5)$$

Where $k_{1i}, k_{2i} \in [-n, -(n-1), \cdots, -1, 0, 1, \cdots, n-1, n]$.

About the number of frequency can be estimated by leveraging the FFT on the subsampled signal, which provides a detailed representation of the frequency components present in the original data. This approach not only enhances the accuracy of frequency estimation but also facilitates the identification of potential noise artifacts that may affect the overall analysis.

## 4. Experiment and Discussion

As mentioned in the introduction, in practical applications, with the continuous expansion of signal bandwidth and the dramatic increase of data access, satisfying Nyquist sampling gradually becomes a great challenge. Therefore, it is of great importance to study how to accurately extract the characteristic parameters of signals under reduced sampling rate conditions and sub-Nyquist sampling conditions. At present, the main sub-Nyquist sampling method that has been widely studied and applied is the compressed perception theory, so it is of practical significance to compare the compressed perception theory with the proposed method. We compare and evaluate the feature parameter estimation accuracy and statistical properties of the proposed method with the compressed perception method under different sample lengths and different signal-to-noise ratios. The feature parameters of the components are still taken from the data in Table 1, the observation matrices of compressed sensing are sampled from the orthogonalized Gaussian random matrices, and the sample reconstruction method is taken from the more stable OMP algorithm. In order to achieve higher frequency resolution, the sample length of the original signal is taken as 2048 points, also to study the effect of sample length on the statistical properties of different algorithms, while the sampling rate of the GEA algorithm is 0.01 times of the maximum component.

We also compare and analyze the extraction accuracy of the two algorithms under different lengths, and the minimum sample length M of the compressed signal to satisfy the reconstruction is obtained based on the following empirical equation (6)[18].

$$M \geq K * lg\left(\frac{N}{K}\right) \qquad (6)$$

Where N is the original signal length and K is the sparsity i.e. the number of components is 10.

The sample length in the test is taken as [2M,4M,16M], and the test results are shown in Fig.2

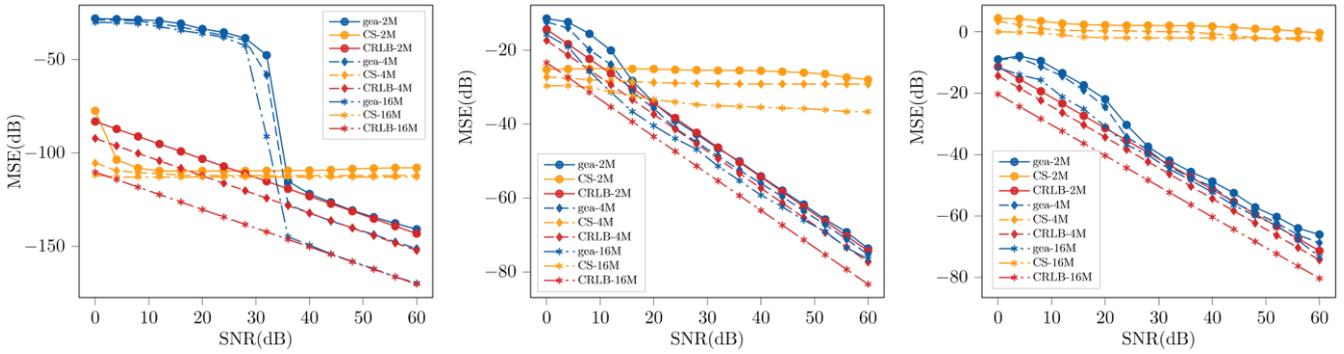

a: frequency estimation    b: amplitude estimation    c: phase estimation

Figure 2

Comparison of sub nyquist's Estimated Performance，M=54

From Fig. 2, the following conclusions can be drawn. When compared with the compressed perception method, the accuracy of the frequency parameters extracted by the proposed method is superior in regions of high signal-to-noise ratio. Furthermore, the proposed method demonstrates an overall advantage in the estimation accuracies of magnitude and phase. However, the CS method is unable to accurately extract the The magnitudes and phases were separated, and it was observed that the estimation accuracies of the phases were very poor. It is important to note that, in the experimental conditions of this paper, the compressed sensing algorithm exhibited a similar tendency to have an obvious constant lower limit with short sample signals. This may be related to compressed sensing in the sampling phase of multi-frequency signals in the signal. The minimum interval for sampling, which is the minimum interval required to determine the corresponding Fourier transform frequency, and the maximum resolution of the frequency are determined. Even with a high signal-to-noise ratio, the FFT is unable to resolve frequencies that exceed the resolution limit, resulting in a persistent estimation error that is independent of the signal-to-noise ratio. Consequently, the estimates are biased. Furthermore, although random sampling preserves the sparse vector nature of the signal by satisfying the constrained isometric (RIP) property, which allows compressed sparse vectors to uniquely recover signals, the process also introduces random aliasing noise, which further degrades the estimation accuracy of the CS. The proposed method, GEA, does not suffer from this limitation due to its distinctive capability of extracting theoretically precise values of the characteristic parameters, which enables it to retain its capacity to estimate the frequency, amplitude, and phase of m.

## 5. Conclusion

The proposed super-resolution generalized eigenvalue method effectively addresses the limitations of existing sub-Nyquist sampling techniques. By leveraging the matrix pencil method and CRT, this approach enhances the accuracy of frequency, amplitude, and phase estimation in high-frequency signal extraction. Experimental results demonstrate that the proposed method outperforms traditional compressed sensing techniques, particularly in high signal-to-noise ratio scenarios. This advancement holds promise for various applications, including telecommunications, radar, and medical imaging, by enabling more accurate and efficient signal processing under reduced sampling rate conditions.